УДК 378.091:004.4

**Величко Владислав Євгенович**
кандидат фізико-математичних наук, доцент, докторант
ДЗ «Луганський національний університет імені Тараса Шевченка», м. Старобільськ, Україна
*vladislav.velichko@gmail.com*

# ВІЛЬНЕ ПРОГРАМНЕ ЗАБЕЗПЕЧЕННЯ В ЕЛЕКТРОННОМУ НАВЧАННІ МАЙБУТНІХ УЧИТЕЛІВ МАТЕМАТИКИ, ФІЗИКИ ТА ІНФОРМАТИКИ

**Анотація.** Популярність використання вільного програмного забезпечення в ІТ-індустрії значно вища, ніж його використання в освітній діяльності. Недоліки вільного програмного забезпечення і проблеми його впровадження в навчальний процес є стримуючим фактором для системного використання його в освіті, однак, відкритість, доступність та функціональність є головними факторами впровадження вільного програмного забезпечення в освітній процес. Утім, для майбутніх учителів математики, фізики та інформатики вільне програмне забезпечення призначене якнайкраще через специфіку його створення, а тому, постає питання системного аналізу можливостей використання вільного програмного забезпечення в електронному навчанні майбутніх учителів математики, фізики та інформатики.

**Ключові слова:** електронне навчання; вільне програмне забезпечення; підготовка вчителів математики, фізики та інформатики.

## 1. ВСТУП

Плідна діяльність особистості, яка включає в себе інформаційну складову і з кожним роком збільшується й розширюється через вдале використання інформаційно-комунікаційних технологій (ІКТ) має бути одним з основних чинників інформатизації освіти. Розширення форм, методів та засобів навчання, які основані на використанні ІКТ спонукають дослідників не тільки до їх використання, а й для системного аналізу їхніх можливостей, очікуваних переваг та можливих недоліків. Сьогодення використання ІКТ в освітній діяльності набуло великого досвіду широкого й експериментального застосування інноваційних технологій. На цей час виконується поєднання різноманітних технологій навчання, основаних на використанні ІКТ, а тому, постає питання розгляду й аналізу отриманих результатів їх використання.

**Постановка проблеми.** Значимим об'єктом комп'ютерної індустрії є вільне і відкрите програмне забезпечення. Започатковане як філософське поняття вільне програмне забезпечення не тільки знайшло своїх прихильників, а й має у своєму арсеналі велику кількість програмних засобів, що застосовуються в різноманітних напрямках. Не є виключенням і навчальна діяльність педагогічних вишів, у яких використовуються системи для організації дистанційної освіти MOODLE, ILIAS; системи комп'ютерної математики MATLAB, Octavia, GAP, SPP; графічні пакети GIMP, Inkscape; офісні пакети Apache Open Office, Libre Office. Однак, залишається питання щодо повноцінного і систематичного використання вільного програмного забезпечення в навчальній діяльності взагалі і в підготовці майбутніх учителів математики, фізики та інформатики зокрема, через те, що серед педагогічних спеціальностей саме вчителі цих предметів мають можливість повною мірою використовувати існуюче вільне і відкрите програмне забезпечення у власній професійній діяльності.

**Аналіз останніх досліджень і публікацій.** Загальні проблеми вільного програмного забезпечення, юридичні і філософські аспекти його існування і використання висвітлюються в роботах Дж. Гослінга, Е. Реймонда, Р. Столлмана та ін. В Україні проблемам використання вільного програмного забезпечення в системі освіти присвятили свої роботи Є. Алексєєв, Г. Злобін, Л. Панченко, С. Семеріков, І. Теплицький, В. Хахановський та ін. Дослідження питання електронного навчання відображені в роботах В. Бикова, Р. Кларк (R. Clark), Р. Майер (R. Mayer), М. Росенберг (M. Rosenberg), М. Шишкіної, С. Щеннікова та ін. Сучасний стан питання підготовки вчителів математики, фізики та інформатики з використанням ІКТ було розкрито в роботах Н. Морзе, М. Жалдака, О. Співаковського, О. Спіріна, Ю. Тріуса та ін.

**Мета статті.** Метою нашого дослідження є аналіз можливостей використання вільного програмного забезпечення в електронному навчанні майбутніх учителів математики, фізики та інформатики відповідно до специфіки їх підготовки.

## 2. РЕЗУЛЬТАТИ ДОСЛІДЖЕННЯ

Ідеї використання механічних систем у начальній діяльності ще на початку XX століття розпочалось у роботах психолога С. Прессей (S. Pressey) розробкою машин для тестування [16]. Фундаментальна праця Б. Скіннер (B. Skinner) "The Science of Learningand Artof Teaching" заклала основи програмованого навчання з лінійним алгоритмом, яке потім було розвинуте й доповнене розгалуженим алгоритмом Н. Кроудер (N. Crowder). Ціла низка як вітчизняних, так і закордонних учених займалась проблемою програмованого навчання з використанням комп'ютерних технологій [1; 9; 17].

На наступному етапі використання ІКТ в навчальній діяльності з'явились мультимедійні компоненти. Саме тоді С. Пейперт (S. Papert), спираючись на дослідження Ж. Піаже (J. Piaget) започаткував використання абстрактного навчального середовища, у якому дитина може створювати і використовувати нові абстрактні поняття. і головне, що для створення цього середовища передбачалось використання персонального комп'ютера.

Сьогодення використання ІКТ в навчальному процесі характеризується, перш за все, додаванням широких можливостей комп'ютерних комунікаційних технологій і побудованих на їх основі хмарних технологій. Використання ІКТ для побудови дистанційної освіти, масових відкритих он-лайн курсів, гібридного навчання і мобільного навчання породжують нові вимоги як до методики використання програмного забезпечення, так і до самого програмного забезпечення [3].

Споріднену історичну градацію наводить С. Семеріков, виділяючи перший етап впровадження ІКТ в освітню діяльність (20–50-ті роки XX сторіччя) і характеризуючи його як період застосування механічних, електромеханічних і електронних індивідуалізованих пристроїв, за допомогою яких подавався навчальний матеріал і виконувався контроль і самоконтроль знань – технологія програмованого навчання. Другий етап (50–80-ті роки XX сторіччя), якому було надано характеристику широкого впровадження ЕОМ у практику. Третій етап (з 80-х років минулого сторіччя) характеризується як етап персональних комп'ютерів і комп'ютерних мереж [5].

Ідеологія вільного програмного забезпечення, а потім і відкритого програмного забезпечення, спираючись на свої визначені ступені свободи декларує навчання на відкритому коді як одну з особливостей вільного програмного забезпечення. Такий підхід є вирішальним фактором для таких напрямків навчання, як «комп'ютерні науки», «програмування», «комунікаційні технології» тощо. Його можна назвати першим етапом використання вільного програмного забезпечення в освітній діяльності, й етапом мікроінформатики і відкритих систем у загальній еволюції ІКТ в освітній діяльності. На цьому етапі в освітній діяльності через відсутність вільного і відкритого програмного забезпечення навчального призначення були використані тільки мови програмування і розглядались розроблені відкриті стандарти взаємодії обчислювальних систем.

Другий етап використання вільного програмного забезпечення пов'язаний із широкими комунікаційними можливостями у створенні розподілених і дистанційних навчальних ресурсів, розвиток яких передусім пов'язаний з розвитком вільного програмного забезпечення. Багато навчальних закладів створюють власні системи дистанційної освіти і системи підтримки освіти на основі ІКТ, що базуються на вільному програмному забезпеченні.

Третій етап характеризується розширенням мультимедійних можливостей комунікаційних технологій і створенням систем керування контентом Web 2.0. На цьому етапі продовжують розвиватись можливості систем дистанційного навчання, з'являється мобільне, відкрите та гібридне або змішане навчання. Особливої уваги, з точки зору вільного програмного забезпечення, слід приділити створенню і розвитку концепції хмарних технологій. Адже, з точки зору вільного програмного забезпечення, що є чудовим варіантом тонкого клієнта з використанням відкритих форматів файлів, є можливість створення принципово нового глобального інформаційно-освітнього відкритого середовища, одним з компонентів якого є вільне програмне забезпечення. Тут уже постає питання про узагальнення методики використання програмного забезпечення, об'єднавши наведені форми навчання в єдиний об'єкт дослідження.

Розв'язання поставленого нами питання стосується, перш за все, аналізу існуючих форм і методів використання ІКТ в навчальному процесі, але воно не може бути прийнятим як остаточне, тому що ґрунтується виключно на всебічному аналізі історичних етапів використання ІКТ через такі міркування. По-перше, дослідження використання ІКТ в навчальному процесі в історичному аспекті не дозволяє виділити вплив основних рушійних сил процесу використання ІКТ. По-друге, виникає проблема прогнозування розвитку системи через відсутність закономірностей і критеріальних показників. По-третє, неможливо визначити інші можливі варіанти розвитку ситуації в ті чи інші вирішальні моменти. А тому розглянемо питання використання вільного програмного забезпечення в електронному навчанні майбутніх учителів математики, фізики та інформатики з точки зору системного підходу.

Термін «електронне навчання», «е-навчання» (Electronic Learning, E-learning) за різними даними було вперше використано в період 1997–1999 років. Його поява пов'язана з іменами Е. Масіе (E. Masie) і Дж. Кросс (JayCross) і був введений не як термін з професійної педагогіки, а як термін з підприємництва і популяризації науки. На це вказує й акцентування уваги на технологічний аспект на противагу педагогічному. Термін Electronic Learning замінив такі терміни як Web (Computer) Based Training, Online Learning та ін. В останніх термінах акцентується увага на технічну платформу або режим доступу. У терміна Electronic Learning основа поняття свідчить на відсутність технічних і організаційних форм з усвідомленням того, що можна використовувати в навчанні будь-які електронні пристрої для обробки інформації. Електронне навчання характеризує не новий вид навчання, а інтегрує в собі низку термінологічних понять у сфері застосування сучасних ІКТ в освіті таких як: комп'ютерні технології навчання, інтерактивні мультимедійні засоби, навчання на основі web-технологій, он-лайн навчання та інші. Попри це, термін «електронне навчання» характеризує окремі види навчальної організації такі, наприклад, як дистанційне навчання. Це пов'язано із застосуванням інформаційно-комунікаційних технологій у сучасних системах дистанційної освіти і з широким упровадженням технологій дистанційної освіти в традиційних університетах. Отже, стираються грані між навчанням на відстані і безпосередньої присутності всередині університетських аудиторій. Саме цю інтеграцію дистанційної і традиційної організації навчального процесу на основі інформаційно-комунікаційних технологій більш адекватно відображає термін «електронне навчання».

В. Биков у наведеному контексті розглядав електронне дистанційне навчання як різновид дистанційного навчання, за яким учасники й організатори навчального процесу здійснюють переважно індивідуалізовану взаємодію як асинхронно, так і синхронно, переважно і принципово використовуючи електронні транспортні системи доставки засобів навчання та інших інформаційних об'єктів, комп'ютерні мережі Інтернет/Інтранет, медіа навчальні засоби та інформаційно-комунікаційні технології [4, с. 191–193].

На думку фахівців, електронне навчання охоплює всі форми навчання і викладання, що відбуваються за допомогою електронної підтримки і спрямовані на формування знань з урахуванням індивідуального досвіду. При цьому електронна підтримка передбачає використання різноманітних інформаційних і комунікаційних систем, будь-то мережеві чи локальні. Так, М. Росенберг (M. Rosenberg) дав таке тлумачення терміну електронне навчання. За його висновками електронне навчання – це використання Інтернет-технологій для надання широкого спектра рішень, які забезпечують підвищення знань і продуктивності праці й базується на роботі в мережі, у якій доставка навчального контенту кінцевому користувачу здійснюється за допомогою комп'ютера [15].

Дослідники Р. Кларк (R. Clark) і Р. Майер (R. Mayer) визначили електронне навчання як навчання у вигляді інструкцій, які поставляються на цифрові пристрої, такі як комп'ютер або мобільний пристрій, що призначені для підтримки навчання. Вони виділяють такі особливості електронного навчання [12]:

– різноманітні способи отримання навчального матеріалу (придбання на компакт-дисках, на засобах збереження інформації, у локальних або хмарних файлових архівах);
– контент навчального матеріалу повністю відповідає об'єкту навчання;
– широке використання мультимедійних елементів у навчальному контенті, використовуючи слова у вигляді усного або друкованого тексту і зображень, таких як ілюстрації, фотографії, анімації або відео;
– використання різноманітних методів навчання: приклади, практичні завдання, зворотній зв'язок;
– може бути під керівництвом інструктора (синхронне електронне навчання) або призначені для самостійного вивчення (асинхронне електронне навчання);
– допомагає тим, хто навчається, будувати нові знання і навички, пов'язані з індивідуальними цілями навчання або підвищення організаційної ефективності.

Дослідники Ф. Майадас (F. Mayadas), Г. Міллер (G. Miller) та Дж. Сенер (J. Sener) у своєму дослідженні запропонували розглядати поняття електронного навчання через сім сучасних моделей його використання [14]. Отже, маємо змогу зробити висновок, що електронне навчання є не що інше, як синтез різноманітних форм і методів навчання з використанням інформаційно-комунікаційних технологій.

Аналізуючи тенденції розвитку систем електронного навчання, окреслюючи сучасні проблеми і протиріччя розвитку систем електронного навчання, таких як доступність навчання; якість освітніх послуг, індивідуалізація навчання, ризики і переваги використання комп'ютерної техніки; стандартизація технологій і ресурсів, М. Шишкіна як платформу електронного навчання запропонувала використовувати хмарні технології [10].

Різноманітність форм і методів електронного навчання визначається не тільки організаційними формами, методиками використання інформаційно-комунікаційних технологій, апаратними ресурсами, а й програмним забезпеченням, завдяки якому й організовуються технологічні процеси. Аналізуючи програмне забезпечення, за допомогою якого організується навчальна діяльність, слід зазначити, що завдяки маркетинговим зусиллям, технічній і методичній підтримці програмних продуктів на першому місці з'являється пропрієтарне програмне забезпечення. Не зменшуючи в жодному разі переваг пропрієтарного програмного забезпечення, слід зазначити, що існує великий клас програмного забезпечення. який, з тих чи інших причин, не використовується масово в навчальній діяльності, хоча його технічні характеристики і навчальні можливості не менш значущі, ніж у пропрієтарного – вільне програмне забезпечення.

Серед стримуючих факторів використання вільного програмного забезпечення в освітній діяльності можна виділити такі: фінансовий (необхідні кошти на навчання викладачів і вчителів), психологічний (існує недовіра до вільного програмного забезпечення), відсутність підтримки (відсутні методики використання вільного програмного забезпечення), несумісність з пропрієтарним програмним забезпеченням (закриті формати файлів, невисока сумісність), відсутність аналогів (для деяких класів програмного забезпечення відсутні аналоги) [2].

Основою електронного навчання, перш за все, є надання освіти за допомогою засобів передавання інформації. До засобів передавання інформації і її подання залучаються такі ресурси, як Інтернет, Інтранет, супутникове та ефірне телебачення, аудіо та відео матеріали на будь-яких носіях. В останні кілька років вищі навчальні заклади ініціювали створення систем з відкритим вихідним кодом. Системи електронного навчання є першими кроками в ініціативі вищої освіти відійти від пропрієтарного програмного забезпечення в сторону вільного програмного забезпечення. Створене програмне забезпечення дозволяє вищим навчальним закладам легко й вільно перевіряти і розповсюджувати свою систему. Система стає відкритою і прозорою, знижуючи залежність від постачальника програмного забезпечення. Також система стає гнучкою через можливість її модернізації і переробки. Попри це, відкрита система дозволяє розширювати обмін ідеями, що, у свою чергу, необхідно для поширення інновацій. Дослідник Дж. Юнг (J. Young) стверджував, що будь-яка людина може використовувати програмне забезпечення з відкритим вихідним кодом, однак, успішна реалізація моделі з відкритим вихідним кодом, залежить від: 1) соціального настрою суспільства, 2) узгодження єдиного визначення відкритого джерела, 3) виділення й закріплення бюджету вільного програмного забезпечення, 4) заохочення організацій, щоб перейти до відкритих вихідних кодів, і 5 ) позитивні робочі відносини з компаніями [18].

У дослідженні К. Коппола (C. Coppola) і Е.Неллі (E. Neelley) окреслені декілька переваг використання відкритого програмного забезпечення, які полягають в такому [13]:

- програмне забезпечення розвивається більш швидко й органічно;
- потреби користувачів дуже близькі до моделі відкритого програмного забезпечення через використання колективного досвіду й особистого внеску розробників;
- нові версії виходять досить часто і тестування, ретельно виконане співтовариством користувачів і розробників, дає змогу в результаті отримати якісне забезпечення тестування на більшій кількості платформ, і у великій кількості середовищ;
- команда розробників часто складається з добровольців, які розташовані один від одного на значних відстанях і є фахівцями в різних сферах діяльності;
- посилюється безпека програмного забезпечення через те, що у відкритому коді кожен із користувачів може знайти його недоліки і, або самостійно їх виправити, або вказати на їх наявність.

Підготовка вчителів математики, фізики та інформатики передбачає, на відміну від учителів інших напрямків, досить суттєву підготовку, спрямовану на формування інформатичної компетентності майбутніх учителів. До комплексу заходів цієї підготовки різні автори відносять різноманітні форми навчання. Так, доречність використання електронного навчання в підготовці майбутніх учителів математики, фізики та інформатики підтверджується в дисертаційній роботі О. Співаковського, який стверджував, що впровадження сучасних інформаційних технологій сприяє більш ефективному опануванню системи знань і вмінь, розвиває творчу спрямованість студента, допомагає формуванню відповідних професійних і особистісних якостей. При цьому сучасні інформаційні технології не самоціль, а педагогічно виправданий підхід, що має розглядатися передусім у плані педагогічних переваг, котрі вони спроможні забезпечити порівняно з традиційною методикою навчання [6].

Беручи до уваги дослідження О. Спіріна ([7]), у яких він зазначив, що специфіка вивчення дисциплін, які формують інформатичну компетентність учителя інформатики, потребує виконання значної кількості лабораторних робіт, які потребують наявності ліцензованого програмного забезпечення, до якого, у першу черг, належить вільне програмне забезпечення. Системи програмування загального призначення Code::Block, Geany, KDevelop, NetBeans та багато інших дозволяють вести розробку програмного забезпечення на багатьох найбільш популярних мовах програмування, системи Dec-C++, Lazarus, Qt-Creator та інші є розвиненими системами для певних мов програмування, навіть існує відкрита система розробки, яка орієнтована на хмарні технології – Codebox.

Окрім програмування у підготовці вчителів математики, фізики та інформатики використовуються такі математичні пакети, як системи комп'ютерної алгебри Maxima і GAP, система математичних обчислень Octave та комп'ютерної математики Scilab, пакети статистичного аналізу R, PSPP та багато інших. У наведеній Ю. Тріусом класифікації комп'ютерно-орієнтованих методів, засобів і форм організації навчання у ВНЗ, не тільки існує вільне програмне забезпечення у наведених компонентах технології навчання, а й використовується на практиці, про що свідчить наявність у запропонованій автором таблиці порівняння найбільш відомих і поширених у світі систем комп'ютерної математики системи Scilab [10].

Результати виконання лабораторних робіт і науково-дослідницької діяльності можуть бути представлені презентативними можливостями пакетів офісних додатків Apache Open Office, Libre Office та інші. При цьому, особливе місце як у науковому світі, так і в підготовці математичної документації займає вільна і відкрита система підготовки тексту TeX і її розширення LaTeX.

Організація системи електронного навчання передбачає використання систем керування навчанням. Безумовним лідером з поширеності систем керування навчанням, серед вільного і відкритого програмного забезпечення, є система Moodle, що відповідає стандарту SCORM і має модульну структуру, завдяки якій для неї створено велику кількість додаткових модулів, що розширяють можливості системи. Серед подібних систем керування навчанням слід розглянути такі системи, як ATutor, Chamilo і eFront. До недоліків кожної з них, а також системи Moodle, слід віднести незадовільну локалізацію. При цьому треба зазначити, що через новітність електронного навчання, і як результат появі нових термінів, не завжди існує переклад на широковживані та сталі терміни. Наступним їх недоліком є недостатня кількість документації, особливо локалізованих, і якщо для системи Moodle виконана достатня кількість наукових досліджень і методичних розробок, то для решти їх не існує. До переваг наведених систем слід віднести направленість на інтеграцію із соціальними мережами і хмарними технологіями, направленість на мобільні клієнти та файлові і відеосховища, направленість на підтримку он-лайн конференцій і сучасних форм навчання. Отже, маємо, з одного боку. розповсюджену систему, побудовану за модульним принципом, для якої розроблені методичні поради щодо її використання, а з іншого, – більш динамічні та мобільні системи, які створені відповідно до сучасних ергономічних розробок.

## 3. ВИСНОВКИ ТА ПЕРСПЕКТИВИ ПОДАЛЬШИХ ДОСЛІДЖЕНЬ

Наведений огляд вільного програмного забезпечення і його застосування у підготовці вчителів математики, фізики та інформатики говорить не тільки про можливість його використання, а й про цілеспрямований і систематичний процес, результат якого полягає в досягненні поставленої мети підготовки майбутніх учителів. Застосування різнопланового програмного забезпечення призводить до формування однієї з ключових здібностей учителя – усвідомлений і обґрунтований вибір методів розв'язання завдань і програмного забезпечення, яке якнайкраще відповідає запропонованим методам.

Розвиток вільного програмного забезпечення є суттєвим ресурсом розвитку інформаційних технологій і науково-технічного процесу, а тому, подальше систематичне вивчення можливостей вільного програмного забезпечення, форм та методів його використання в освітній діяльності є перспективним завданням дослідження.

## СПИСОК ВИКОРИСТАНИХ ДЖЕРЕЛ

1. Беспалько В. П. Программированное обучение. Дидактические основы / В. П. Беспалько. – М. : Высшая школа, 1970. – 300 с.
2. Величко В. Е. Сдерживающие факторы использования свободного программного обеспечения в университетском образовании / В. Е. Величко //Интеграция науки и практики: проблемы и перспективы развития : материалы II Всероссийской научно-практическойк онференции с международным участием. Старый Оскол, 9−10 апреля 2014 г. / под. ред.: Юриной Н. В., Лаенко Л. В., Рудакова А. В. и др. – Старый Оскол : ООО «Оскольская типография», 2014. – С. 185–188.
3. Воронкін О. С. Періодизація розвитку інформаційно-комунікаційних технологій навчання / О. С. Воронкін // Вища освіта України. –2014. – № 3 (54). – С. 109–116.
4. Енциклопедія освіти / Акад. пед. наук України; головний ред. В. Г. Кремень. – К. Юрінком Інтер, 2008. – 1040 с.
5. Семеріков С. О. Теоретико-методичні основи фундаменталізації навчання інформатичних дисциплін у вищих навчальних закладах : дис... д-ра пед. наук : 13.00.02 / Семеріков Сергій Олексійович ; Національний педагогічний ун-т ім. М. П. Драгоманова. – К., 2009. – 536 с.
6. Співаковський О. В. Теоретико-методичні основи навчання вищої математики майбутніх вчителів математики з використанням інформаційних технологій : автореф. дис... докт. пед. наук : 13.00.02. / О. В. Співаковський ; Нац пед. ун-т ім. М. П. Драгоманова. – Київ, 2004. – 44 с.
7. Спірін О. М. Методична система базової підготовки вчителя інформатики за кредитно-модульною технологією : монографія / Олег Михайлович Спірін. – Житомир : Вид-во ЖДУ ім. І. Франка, 2013. – 182 с.
8. Стратегія розвитку інформаційного суспільства в Україні [Електронний ресурс] / Кабінет Міністрів України. – 2013. – Режим доступу до ресурсу: http://zakon5.rada.gov.ua/laws/show/386-2013-%D1%80
9. Томас К., ДевисДж., Опеншоу Д., БёрдДж. Перспективы программированного обучения : Пер. с англ. – М. : Мир, 1966. – 391 с.
10. Триус Ю. В. Комп'ютерно-орієнтовані методичні системи навчання математичних дисциплін у ВНЗ: проблеми, стан і перспективи / Ю. В. Триус // Науковий часопис НПУ імені М. П. Драгоманова. Серія 2: Комп'ютерно-орієнтовані системи навчання. – 2010. – № 9. – С. 16–29.
11. Шишкіна М. Перспективні технології розвитку систем електронного навчання / М. Шишкіна // Інформаційні технології в освіті.–2011. – № 10. – С. 132–139.
12. Clark R. C., Mayer R. E. E-learning and the science of instruction: Proven guidelines for consumers and designers of multimedia learning. – JohnWiley&Sons, 2011.
13. Coppola C. Open Source – OpenLearning: why open source makes sense for education / C. Coppola, E. Nelly // Presentedat Open Source Summit, 2004 [Електронний ресурс]. – Режим доступу : http://arizona.openrepository.com/arizona/handle/10150/106028. – Назва з екрана.
14. Mayadas F. Definitionsof E-Learning Courses and Programs Version 2.0 April 4, 2015 : Updated E-LearningDefinitions / FrankMayadas, GaryMiller, JohnSener // OLC Insights. –[Електронний ресурс]. – Режим доступу: http://onlinelearningconsortium.org/updated-e-learning-definitions-2/. – Назва з екрана.

15. Rosenberg M. E-Learning: Strategies for Delivering Knowledgein the Digital Age / Marc Rosenberg. – New-York: TheMcGrawHillCompanies, 2001. – 344 p.
16. Pressey S. L. A simple apparatus which gives tests and scores and teaches / S. L. Pressey // SchoolandSociety. – 1926. – № 23. – P. 373–376.
17. Skinner B. F. The science of learning and art of teaching / B. F. Skinner // Harward Education Review. – 1954. – № 24. – P. 86-97.
18. Young J. Five challenges for open source / J. Young //ChronicleofHigherEducation (September, 2004). [Електронний ресурс]. – Режим доступу : http://chronicle.com/article/5-Challenges-for-Open-Source/18715/. – Назва з екрана.

# СВОБОДНОЕ ПРОГРАММНОЕ ОБЕСПЕЧЕНИЕ В ЭЛЕКТРОННОМ ОБУЧЕНИИ УЧИТЕЛЕЙ МАТЕМАТИКИ, ФИЗИКИ И ИНФОРМАТИКИ

**Величко Владислав Евгеньевич**
кандидат физико-математических наук, доцент, докторант
ГУ «Луганский национальный университет имени Тараса Шевченко», г. Старобельск, Украина
*vladislav.velichko@gmail.com*

**Аннотация.** Популярность использования свободного программного обеспечения в ИТ-индустрии значительно выше, чем его популярность использования в образовательной деятельности. Недостатки свободного программного обеспечения и проблемы его внедрения в учебный процесс являются сдерживающим фактором для системного использования его в образовании, однако, открытость, доступность и функциональность являются главными факторами внедрения свободного программного обеспечения в образовательный процесс. Тем не менее, для будущих учителей математики, физики и информатики свободное программное обеспечение предназначено как нельзя лучше из-за специфики его создания, а потому, возникает вопрос системного анализа возможностей использования свободного программного обеспечения в электронном обучении будущих учителей математики, физики и информатики.

**Ключевые слова:** электронное обеспечение; свободное программное обеспечение; подготовка учителей математики, физики и информатики.

# FREE SOFTWARE IN ELECTRONIC LEARNING FUTURE TEACHERS OF MATHEMATICS, PHYSICS AND COMPUTER SCIENCE

**Vladyslav Ye. Velychko**
PhD (in Physical and Mathematical Sciences), associate professor, doctoral candidate
Luhansk Taras Shevchenko National University, Starobilsk, Ukraine
*vladislav.velichko@gmail.com*

**Abstract.** Popularity of the use of free software in the IT industry is much higher than its popular use in educational activities. Disadvantages of free software and problems of its implementation in the educational process is a limiting factor for its use in the education system, however, openness, accessibility and functionality are the main factors for the introduction of free software in the educational process. Nevertheless, for future teachers of mathematics, physics and informatics free software is designed as well as possible because of the specificity of its creation, and therefore, there is a question of the system analysis of the possibilities of using open source software in e-learning for future teachers of mathematics, physics and computer science.

**Keywords:** e-learning; free software; the training of teachers of mathematics, physics and computer science.

## REFERENCES (TRANSLATED AND TRANSLITERATED)

1. Bespal'ko V. P. Programmed learning. Didactic bases. / V. P. Bespal'ko. – M. :Vysshajashkola, 1970. – 300 s. (inRussian)
2. Velichko V. E. Constraints of using free software in the university education / V. E. Velichko //Integracija nauki I praktiki: problem I perspektivy razvitija: materialy II Vserossijskoj nauchno-prakticheskoj konferencii s mezhdunarodnym uchastiem. Staryj Oskol, 9-10 aprelja 2014 g. / pod. red.: Jurinoj N. V., Laenko L. V., Rudakova A. V. i dr. – StaryjOskol: OOO «Oskol'skajatipografija», 2014. – S. 185-188. (in Russian)
3. Voronkin O. S. Periods of development ICT training / O. S. Voronkin // Vyshha osvita Ukrajiny, 2014. – # 3 (54). – S. 109–116. (in Ukrainian)
4. Encyklopedija osvity / Akad. ped. Nauk Ukrajiny; gholovnyj red. V. Gh. Kremenj. – K. Jurinkom Inter, 2008. – 1040 s. (in Ukrainian)
5. Semerikov S. O. Theoretical and methodological foundations fundamentalization study computer science in higher education: dys... d-raped. nauk : 13.00.02 – teorija ta metodyka navchannja (informatyka) / Semerikov Serghij Oleksijovych ; Nacionaljny j pedaghoghichnyj un-t im. M. P. Draghomanova. – K., 2009. – 536 s. (in Ukrainian)
6. Spivakovsjkyj O. V. Theoretical and methodical grounds of higher mathematics for future teachers of mathematics using information technology: avtoref. dys... dokt. ped. nauk : 13.00.02. / Oleksandr Volodymyrovych Spivakovsjkyj ;Nacped. un-tim. M. P. Draghomanova. – Kyjiv, 2004. – 44 s. (in Ukrainian)
7. Spirin O. M. Methodical system of training teachers of basic science for credit-modular technology: monograph / Olegh Mykhajlovych Spirin. – Zhytomyr :Vyd-voZhDUim. I. Franka, 2013. – 182 s. (in Ukrainian)
8. Strategy of information society development in Ukraine [online] / Cabinet of Ministers of Ukraine. – 2013. – Available from: http://zakon5.rada.gov.ua/laws/show/386-2013-%D1%80 (in Ukrainian)
9. Tomas K., Devis Dzh., Openshou D., Bjord Dzh. Prospects of programmed training: Per. s angl. – M. : Mir. – 1966. – 391s. (in Russian)
10. TryusJu. V. Computer-oriented methodological training system of mathematical sciences in universities: problems and prospects / Ju. V. Tryus // Naukovyjchasopys NPU imeni M. P. Draghomanova. Serija 2: Komp'juterno-orijentovani systemy navchannja. – 2010. – # 9. – S. 16-29. (in Ukrainian)
11. Shyshkina M. Advanced technologies of e-learning systems / M. Shyshkina // Informacijni tekhnologhiji v osviti. – 2011. – # 10. – S. 132–-139. (in Ukrainian)
12. Clark R. C., Mayer R. E. E-learning and the science of instruction: Proven guidelines for consumers and designers of multimedia learning. – John Wiley & Sons, 2011.(in English).
13. Coppola C. Open Source – Open Learning: why open source makes sense for education [online] / C. Coppola, E. Nelly // Presented at Open Source Summit, 2004. – Available from: http://arizona.openrepository.com/arizona/handle/10150/106028.(in English).
14. Mayadas F. Definitions of E-Learning Courses and Programs Version 2.0 April 4, 2015 : Updated E-Learning Definitions [online] / Frank Mayadas, Gary Miller, John Sener // OLC Insights. — Available from: http://onlinelearningconsortium.org/updated-e-learning-definitions-2/.(in English).
15. Rosenberg M. E-Learning: Strategies for Delivering Knowledge in the Digital Age / Marc Rosenberg. – New-York: The McGraw Hill Companies, 2001. – 344 p. .(in English).
16. Pressey S. L. A simple apparatus which gives tests and scores and teaches / S. L. Pressey // School and Society. – 1926. – № 23. – P. 373–376. .(in English).
17. Skinner B. F. The science of learning and art of teaching / B. F. Skinner // Harward Education Review. – 1954. – № 24. – P. 86–97. .(in English).
18. Young J. Five challenges for open source [online] / J. Young //Chronicle of Higher Education (September, 2004). — Available from : http://chronicle.com/article/5-Challenges-for-Open-Source/18715/.(in English).

**Conflict of interest.** The author has declared no conflict of interest.